\documentclass[prb,aps,twocolumn,superscriptaddress,showpacs]{revtex4}
\usepackage{amssymb}
\usepackage{epsfig}
\usepackage{graphicx}
\usepackage{dcolumn}
\usepackage{bm}
\usepackage{curves}
\usepackage{epic}
\usepackage{wasysym}
\usepackage{subfigure}

\begin{document}

\title{      Optical conductivity due to orbital polarons
             in systems with orbital degeneracy }

\author{     Piotr Wr\'obel}
\affiliation{Max Planck Institute for the Physics of Complex Systems,
             N\"othnitzerstr. 38, D-01187 Dresden, Germany}
\affiliation{Institute for Low Temperature and Structure Research,
             P.O. Box 1410, PL-50950 Wroc{\l}aw 2, Poland}

\author{     Robert Eder}
\affiliation{Karlsruhe Institute of Technology, Institut f\"{u}r Festk\"{o}rphysik,
             D-76021 Karlsruhe, Germany}

\author{     Andrzej M. Ole\'s}
\affiliation{Max-Planck-Institut f\"{u}r Festk\"{o}rperforschung,
             Heisenbergstrasse 1, D-70569 Stuttgart, Germany}
\affiliation{Marian Smoluchowski Institute of Physics, Jagellonian University,
             Reymonta 4, PL-30059 Krak\'ow, Poland}

\date{ \today }

\begin{abstract}
We consider the impact of orbital polarons in doped orbitally ordered
systems on optical conductivity using the simplest generic model
capturing the directional nature of either $t_{2g}$ (or $e_g$) orbital
states in certain transition metal oxides, or $p$ orbital states of cold
atoms in optical lattices. The origin of the optical transitions is
analyzed in detail and we demonstrate that the optical spectra:
(i)~are determined by the string picture, i.e., flipped orbitals along
the hole hopping path, and
(ii)~consist of three narrow peaks which stem from distinct excitations.
They occur within the Mott-Hubbard gap similar to the superconducting
cuprates but indicate hole confinement, in contrast to the spin $t$-$J$ 
model. Finally, we point out how to use the point group symmetry to
classify the optical transitions.
\\
{\it Published in: Phys. Rev. B {\bf 86}, 064415 (2012).}
\end{abstract}

\pacs{78.20.Bh, 71.10.Fd, 75.25.Dk, 75.30.Et}

\maketitle

\section{Introduction}
\label{sec:intro}

Electronic excitations at finite frequency are a common feature of doped
Mott or charge-transfer insulators. Such excitations were observed
experimentally in high temperature superconductors by optical absorption
$\sigma(\omega)$ shortly after these systems were discovered,
\cite{Uch91} and were extensively studied in the $t$-$J$ model by 
several groups,\cite{Mor90,Ste90,Poi93,Jak94,Ede95} providing valuable
insights into the charge dynamics in doped cuprates. It was also 
realized that considerable transfers of spectral weight in the optical 
spectra occur and new states arise which generate spectral intensity within 
the Mott-Hubbard gap.
\cite{Mei91,Esk94,Phi10} These studies have shown that the key
assumption of the Fermi liquid theory that the low-energy excitation
spectrum stands in a one-to-one correspondence with that of a
non-interacting system has to be revised when the electrons interact
strongly. For instance, drastic deviations from the Fermi liquid picture 
are obtained, in the normal state of the copper-oxide high-temperature
superconductors, highlighted by a pseudogap, broad spectral features, 
and the resistivity which increases linearly with temperature.
\cite{Phi09} It was recently established that the optical conductivity 
for CuO$_2$ planes of high temperature superconductors exhibits a 
mid-infrared peak at low doping that gradually develops to a band under 
increasing doping and causes an insulator-to-metal transition.\cite{Nic11}

Optical conductivity studies play also a very important role in other
correlated materials, including Mott insulators with active orbital
degrees of freedom. These systems exhibit rather complex behavior due to 
the interrelation between spin, orbital and charge degrees of freedom. 
Recently it was pointed out that the excitations to $3z^2-r^2$ orbitals 
in high-$T_c$ cuprates are responsible for the observed optical 
conductivity in the insulating state.\cite{Mil11} A study of the 
undoped three-orbital Hubbard model explains the anisotropy of the 
optical conductivity of a pnictide superconductor.\cite{Dag11} Finally,
a well known example in this family of compounds are colossal 
magnetoresitance manganites where their complexity manifests itself in 
the large number of competing magnetic phases in the phase diagrams of 
perovskite or layered materials,\cite{Tok06} including the 
charge-ordered phase at 50\% doping with the optical conductivity 
determined by the $3z^2-r^2$-like occupied orbital which coexists with 
charge order.\cite{Sol01} In undoped LaMnO$_3$ the optical conductivity
$\sigma(\omega)$ shows several features at higher energy,\cite{Kov10}
and their spectral weights change with temperature. These features
and the thermal evolution of their spectral weights
may be well understood by employing a general relation between the
spin-orbital superexchange and the spectral weight distribution in the
optical spectra of Mott insulators.\cite{Kha04} This theory is also 
successful in analyzing the temperature dependence of the low-energy 
optical spectra for high spin excitations in LaVO$_3$,\cite{Miy02} 
where spin-orbital entanglement plays a role.\cite{Ole12}

Detailed investigations of the optical conductivity of the ferromagnetic
(FM) metallic La$_{1-x}$Sr$_x$MnO$_3$ have shown:\cite{Oki95}
(i) a pseudogap in $\sigma(\omega)$ for temperatures above the Curie
temperature $T_C$,
(ii) the growth of the broad incoherent spectrum at low energy
$0<\omega<1.0$ eV under decreasing temperature below $T_C$, and
(iii) a narrow Drude peak.
In these compounds electronic correlations among $e_g$ electrons are
strong and a orbital liquid stabilizes the FM metallic phase.\cite{Fei05}
These features have been successfully explained in the theory which
focuses on the orbital dynamics in a situation when spins do not
contribute and may be neglected.\cite{Hor99} In this situation
orbital polarons\cite{Kil99} determine the transport properties and the
optical spectra. Recently a two-peak structure of the optical
conductivity was discussed for the metallic phase of FM manganites,
\cite{Pak11} with a far-infrared Drude peak accompanied by a broad
mid-infrared polaron peak. It is intriguing whether similar phenomena
occur in other orbital systems as well.

In this paper we investigate the optical conductivity in the
two-dimensional (2D) orbital model with two active $t_{2g}$ orbital
flavors. This model reveals crucial properties we are interested in, and
is applicable either to transition metal oxides with FM planes and active
$t_{2g}$ orbitals when the tetragonal crystal field splits off the $xy$
orbital from the $\{yz,zx\}$ doublet filled by one electron at each site,
as for instance in Sr$_2$VO$_4$,\cite{Mat05}
or to certain $e_g$ planar materials such as K$_2$CuF$_4$ or
Cs$_2$AgF$_4$,\cite{Hid83,Wu07}
or to cold-atom systems
\cite{Jak05} with active $p$ orbitals.\cite{Lu09}
Orbital superexchange which arises in the strongly correlated regime
is responsible for alternating orbital (AO) order at half filling. This
is then the reference state playing a role of the physical vacuum below 
when we consider the optical conductivity for a doped Mott insulator.

The paper is organized as follows. In Sec. \ref{sec:model} we introduce
the microscopic model and specify typical parameters. The model is
solved first for a single hole in Sec. \ref{sec:hole}, where we analyze
the processes of possible hole propagation and the role of string states
in the optical conductivity. Next we present a numerical solution for
the optical conductivity in Sec. \ref{sec:num} and interpret the results
in Sec. \ref{sec:picture}. A short summary and final conclusions are
presented in Sec. \ref{sec:summa}. More technical details on the
performed calculations and on the origin of the optical transitions
that could be of interest only for some readers
are given in the Appendix \ref{sec:appa}.

\section{Generalized orbital $t$-$J$ model }
\label{sec:model}

For the purpose of discussing the optical conductivity in orbitally
degenerate systems we concentrate on the recently introduced
strong-coupling version of the two-orbital Hubbard model for
spinless fermions on the square lattice (when the spins form a FM order
and can be neglected).\cite{Woh08} To be specific, the physical
problem to which our analysis applies is the FM plane of a Mott
insulator with AO order of $t_{2g}$ (or $e_g$ or $p$) orbitals,
i.e., interacting spinless fermions which undergo one-dimensional (1D)
nearest neighbor (NN) hopping with conserved orbital flavor:
\begin{equation}
\label{Ht2g} {\cal H}= 
-t\sum_{\{ij\}{\parallel}b}c^{\dagger}_{i,a}c^{}_{j,a} 
-t\sum_{\{ij\}{\parallel}a}c^{\dagger}_{i,b}c^{}_{j,b} 
+U \sum_i n_{ia}n_{ib}\;.
\label{rawh}
\end{equation}
Here $c^{\dagger}_{i,a}$ and $c^{\dagger}_{i,b}$ are creation
operators for electrons with two orbital flavors, and we consider
$t_{2g}$ orbitals,\cite{Woh08}
\begin{equation}
|a\rangle\equiv |yz\rangle,  \hskip .7cm |b\rangle\equiv |xz\rangle,
\end{equation}
labeled by the index of a cubic axis which prohibits the electron
hopping by symmetry; this notation was introduced for titanium and
vanadium perovskites.\cite{Kha00} The summations in Eq. (\ref{Ht2g})
are carried over pairs $\{ij\}$ of NN sites in the $ab$ plane.

In the following we consider an effective Hamiltonian obtained by
a unitary transformation ${\cal U}$ which eliminates the part of the
Hamiltonian that creates/annihilates double occupancies.
The purpose is not to eliminate energetically costly double occupancies,
but rather to assess accurately to what extent they contribute to the
ground state energy and wave functions.\cite{Woh08}
The effective Hamiltonian reads
\begin{equation}
\label{canon}
{\cal H}_{t_{2g}}={\cal P}  {\cal U}^{-1}{\cal H}{\cal U P},
\end{equation}
where ${\cal P}$ is a projection operator that projects the transformed
Hamiltonian on the low energy Hilbert space and removes all states with
doubly occupied sites. The standard perturbation theory
\cite{Cha77,Wro*2} gives the following expression for the generator
in Eq. (\ref{canon}):
\begin{eqnarray}
{\cal U}&=&1+\frac{t}{U}\Big\{\sum_{
\{ij\}{\parallel}b} (c^{\dagger}_{i,a} n_{ib}c^{}_{j,a} -
c^{\dagger }_{j,a}n_{ib} c^{}_{i,a})\nonumber \\
&+& \sum_{\{ij\}{\parallel}a}(c^{\dagger}_{i,b} n_{ia}c^{}_{j,b}
- c^{\dagger}_{j,b} n_{ia} c^{}_{i,b})\Big\}
+O\left(\frac{t^2}{U^2}\right).
\label{canonical}
\end{eqnarray}

Following this scheme we obtain the generalized orbital $t$-$J$ model
with orbital superexchange $J=4t^2/U$ and three-site effective next
nearest neighbor (NNN) hopping $\tau=t^2/U$ (both expressions apply
when $U\gg t$):\cite{noteu}
\begin{eqnarray}
\label{t2gmodel} {\cal H}_{t_{2g}} &=&{\cal P} \left( {\cal
H}_{t}+{\cal H}_{J} +{\cal H}_{\rm 3s}^{(l)} +{\cal H}_{\rm
3s}^{(d)}\right){\cal P}\,,\\
\label{Ht} {\cal H}_{t}&=&-t \sum_{i}
\left({c}^{\dagger}_{i,b}{c}^{}_{i+{\bf \hat{a}},b} +
{c}^{\dagger}_{i,a}{c}^{}_{i+{\bf \hat{b}},a}+\mbox{H.c.}\right)\,,\\
\label{HJ} {\cal H}_{J}&=& \frac12 J \sum_{\langle ij\rangle }
\left(T^z_i T^z_j - \frac{1}{4}{n}_i{n}_j\right)\,,\\
\label{H3sl} {\cal H}_{\rm 3s}^{(l)} &=& -\tau \sum_{i}
\left({c}^\dag_{i-{\bf \hat{a}},b}{n}^{}_{i,a}
{c}^{}_{i+{\bf \hat{a}},b}+\mbox{H.c.}\right) \nonumber\\
&-&\tau\sum_{i}\left({c}^\dag_{i-{\bf\hat{b}},a}
{n}^{}_{i,b}{c}^{}_{i+{\bf \hat{b}},a}+\mbox{H.c.}\right)\,,\\
\label{H3sd}
H_{\rm 3s}^{(d)} &=&-\tau \sum_{i}
\left({c}^\dag_{i\pm{\bf \hat{b}},a}{c}^{}_{i,a}
{c}^\dag_{i,b}{c}^{}_{i\pm{\bf \hat{a}},b}+\mbox{H.c.}\right) \nonumber\\
&-&\tau \sum_{i} \left({c}^\dag_{i\mp{\bf \hat{b}},a}{c}^{}_{i,a}
{c}^\dag_{i,b}{c}^{}_{i\pm{\bf \hat{a}},b}+\mbox{H.c.}\right)\,.
\end{eqnarray}
Here the summations are carried over sites $i\in ab$ plane, and the unit
vectors $\{{\bf \hat{a}},{\bf \hat{b}}\}$ indicate the bond direction in
the $ab$ plane. The superexchange $J$ is Ising-like and couples NN 
orbital operators,
\begin{equation}
T^z_i=\frac{1}{2}\left({n}_{ia}-{n}_{ib}\right),
\end{equation}
on the bonds $\langle ij\rangle$ in the $ab$ plane. The NN hopping $t$
and the effective NNN hopping $\tau$ contribute only in presence of
holes as the projection operators ${\cal P}$ project onto the subspace 
without double occupancies; for more details see Ref. \onlinecite{Woh08}. 
As we demonstrate below, a nice feature of the Hamiltonian 
(\ref{t2gmodel}) is that even on the infinite lattice it can be in 
principle exactly solved in the low energy sector by numerical methods.

\section{A single hole problem}
\label{sec:hole}

\begin{figure}[t!]
 \centering
\includegraphics[width=8.4cm]{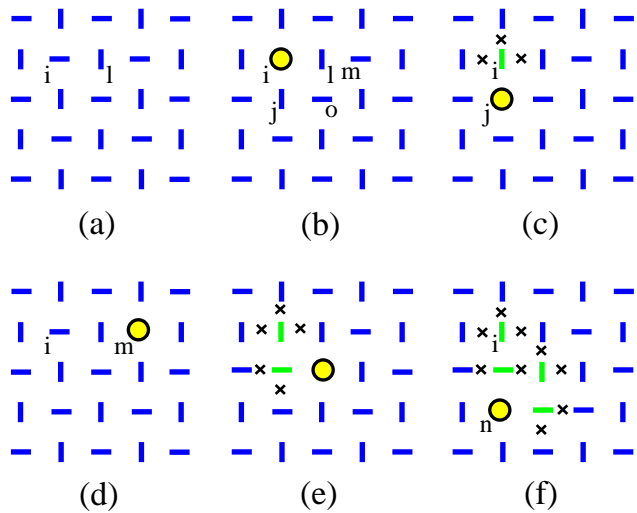}
\caption{(Color online) Artist's view of hole propagation in the ground
state $|{\rm AO}\rangle$ with AO order presented in (a), with vertical
(horizontal) bars standing for $a$ ($b$) occupied orbitals.
(b) A hole doped  into the physical vacuum $|{\rm AO}\rangle$ at site $i$
removes four exchange bonds. It can be shifted to new positions by
consecutively applying different terms of the Hamiltonian:
(c) a single NN hopping $t$,
(d) a single three-site term $\tau$ (\ref{H3sl}),
(e) a single three-site term $\tau$ (\ref{H3sd}), and
(f) several steps which create more flipped orbitals;
for more details see text. Broken bonds
that cost the energy $\frac12 J$ each are marked by~$\times$.}
\label{strings}
\end{figure}

We start the analysis of the optical conductivity by considering the
problem of a single hole. In a Mott insulator when there is exactly one 
electron per lattice site only the exchange term Eq. (\ref{HJ}) 
contributes and the ground state is the `orbital N\'eel state' 
$|{\rm AO}\rangle$ with AO order, playing here a role of the physical 
vacuum and shown schematically in Fig. \ref{strings}(a). Boxes aligned 
along the $\hat\textbf{a}$ ($\hat\textbf{b}$) lattice direction 
represent $b$ ($a$) orbitals, the sublattices containing  $a$ and $b$ 
orbitals in the state $|{\rm AO}\rangle$ will be denoted by $\cal A$ 
and $\cal B$, respectively.
As interactions are Ising-like, quantum fluctuations are absent and
the energy of this state is exactly $-\frac14 J$ per bond. This energy
plays a role of the reference energy of the physical vacuum state 
$|{\rm AO}\rangle$ in what follows.

Next we assume that an electron is removed from the $|{\rm AO}\rangle$
state, i.e., a single hole is created in the state with AO order, see 
Fig. \ref{strings}(b). It will be seen that the motion of this hole 
disrupts the AO order so that the
problem has a strong similarity with the much studied problem of
hole motion in an Ising antiferromagnet, sometimes referred to as the
$t$-$J_z$ model or, more generally, hole motion in an antiferromagnet.
\cite{Poi93,Brinkman_Rice,Trugman,Inoue_Maekawa,Voijta_Becker}
There are two differences:
(i) the hole motion is directional, i.e., a hole on the $\cal A$ 
sublattice can move only in $\hat\textbf{b}$-direction and vice versa, 
and
(ii) the term ${\cal H}_{\rm 3s}^{(l)}$ which is usually neglected in 
the spin $t$-$J$ model but is the only term here responsible for 
coherent motion of the hole.

We consider the single-hole state $c _{ib}|{\rm AO}\rangle$ shown in 
Fig. \ref{strings}(b). In the following we refer to the states shown in 
Fig. \ref{strings} by their labels (b), (c), {\it etcetera}. Creation 
of the hole at site $i$ raises the expectation value of ${\cal H}_J$ 
by $\Delta E_0=\frac14 zJ$, where $z$ is the number of NNs (here $z=4$).
The term ${\cal H}_t$ --- which in principle has the largest matrix 
element $t$ --- couples the state (b) with the state (c). Thereby a 
misaligned orbital is created at $i$, which further increases the 
expectation value of ${\cal H}_J$ by $\Delta E_1=\frac14 (z-1)J$. The 
same holds true for all subsequent hops which involve ${\cal H}_t$ --- 
these create a `string' of misaligned orbitals, see the states (e) and 
(f). Each of these further misaligned orbitals created in step $n>1$ 
increases the energy by $\Delta E_n=\frac14 (z-2)J$. The term 
${\cal H}_t$ therefore does not lead to a coherent propagation of the 
hole. Trugman has discussed coherent hole motion in the $t$-$J_z$ model 
whereby a hole performs one and a half circular movement along the 
smallest closed loop, i.e., around a plaquette in a square lattice.
\cite{Trugman} It is straightforward to see that for this smallest  
$2\times 2$ loop this mechanism does not work in the present case due 
to the directional nature of the hopping ${\cal H}_t$ Eq. (\ref{Ht}) 
that excludes the hopping $t$ along closed loops.

In contrast to this, the term (\ref{H3sl}) couples the states (b) and (d)
without creating any defect in the orbital order. This term therefore
enables true coherent motion
in the insulating ground state with AO order.\cite{Woh08}
Finally, the term (\ref{H3sd}) has yet another
effect in that it connects the states (b) and (e),
as well as the states (e) and (f). In other words this term connects
string states whose number of defects differs by two.
In some special cases the term (\ref{H3sl}) may `split off' clusters of
misaligned orbitals. Consider for example the state (f) and assume
that the hole moves upward to site $i$ by virtue of hopping $\tau$ given
by Eq. (\ref{H3sl}). Then a $2\times 2$ cluster of misaligned orbitals,
all inverted with respect to the AO order, remains next to the hole
in the final state.

In order to discuss the hole motion we restrict the Hilbert space to
a basis of string states
\cite{Brinkman_Rice,Trugman,Inoue_Maekawa,Voijta_Becker} which are
created by successive application of ${\cal H}_t$ starting from the 
state (b). Unlike the case of a quantum antiferromagnet, the Hamiltonian
(\ref{t2gmodel}) does not produce quantum fluctuations of the orbital
order so that this restriction represents an even better approximation
than in the spin $t$-$J$ model. Due to the directional nature of 
${\cal H}_t$ and the orbital N\'eel order each hop along the string must 
be perpendicular to the preceding one so that the maximum number of
different strings created after $n$ hops is $2^n$.
The actual number of topologically different strings is less than this
because one has to exclude self-intersecting paths. If we denote the
sequence of sites visited by the hole by ${\cal F}=\{ i_0,i_1,\dots i_n \}$
and introduce the `orbital-flip operator' at site $i$,
\begin{equation}
\tilde{S}_{i} = c^{\dag}_{i,\bar{o}(i)} \;c^{}_{i,o(i)},
\end{equation}
where $o(i)$ ($\bar{o}(i)$)
denotes the orbital at site $i$ which is occupied (unoccupied)
in the reference $|{\rm AO}\rangle$ state,
the corresponding string state can be written as
\begin{equation}
\label{string}
|\Psi_{i,{\cal F}} \rangle=c_{i_n,o(i_n)} \;
\prod_{j \in {\cal F}'}\;
\tilde{S}_{j}|{\rm AO}\rangle,
\end{equation}
where it is understood that $i=i_0$ and
${\cal F}={\cal F}'\cup\{i_n\}$.
Since we want to study coherent hole motion, we construct Bloch states
out of string states (\ref{string}),
\begin{eqnarray}
\left|\Psi_{{\bf k},{\cal S},{\cal F}} \right\rangle &=&
\sqrt{\frac{2}{N}}\sum_{j}\;e^{i {\bf k} {\bf R}_j}\;
T_{{\bf R}_j} |\Psi_{i,{\cal F}} \rangle,  \\
\left|\Psi^{(n)}_{{\cal S},{\bf k}}\right\rangle&=&
\sum_{{\cal F}} \;\alpha^{(n)}_{{\cal S},{\cal F},{\bf k}}
|\Psi_{{\bf k},{\cal S},{\cal F}} \rangle.
\label{psin}
\end{eqnarray}
Here we have introduced an additional sublattice index
${\cal S}\in \{ {\cal{A}},{\cal{B}} \}$, whereby it is understood that
$i\in {\cal S}$ and the sum over $j$ extends all translations of one
sublattice. The coefficients 
$\{\alpha^{(n)}_{{\cal S},{\cal F},{\bf k}}\}$ are variational 
parameters and $n$ stands for a band index. An analogous {\em ansatz} 
for the related problem of a hole in a quantum antiferromagnet was used 
before in Refs. \onlinecite{Trugman,Inoue_Maekawa,Voijta_Becker}.
In practice, all different sets ${\cal F}$ up to a maximum
number $n_{max}$ of defects are generated by computer
and the Hamiltonian matrix is set up.

We have used $n_{max}=10$ and verified that the results for
the low energy bands are well converged with respect to $n_{max}$.
The matrix elements of the Hamiltonian for the pairs of states
$\{$(b),(c)$\}$, $\{$(b),(d)$\}$ and $\{$(b),(e)$\}$ are $t$, $\tau$,
and $-\tau$ respectively. The sign change with respect to the
Hamiltonian (\ref{t2gmodel}) follows from the transformation of hopping
terms to the hole picture. Due to the directional nature of
the hopping terms the resulting band structure depends on the sublattice
index ${\cal S}$.

\begin{figure}[t!]
 \begin{center}
\includegraphics[width=8.4cm]{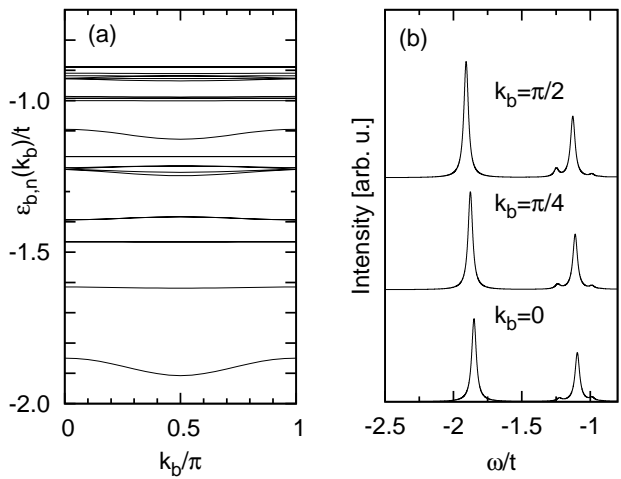}
\end{center}
\caption{ Numerical results obtained for 20
bands with lowest energies, formed by propagating eigenstates 
$|\Psi^{(n)}_{{\cal B},{\bf k}}\rangle$:
(a) energy dispersions along the (0,1) direction in the 2D 
Brillouin zone, and
(b) the spectral function (\ref{Ak}) for representative values of 
$k_b$. Parameters: $J=0.4 t$, $\tau=0.1 t$, $\delta=0.02t$.}
\label{bands}
\end{figure}

Fig. \ref{bands}(a) shows the dispersion of the lowest bands, 
$\epsilon_{b,n}({\bf k})$ as a function of $k_b$, where ${\bf k}=(0,k_b)$
is the vector from the 2D Brillouin zone. The bands are obtained for hole 
doping at ${\cal B}$ sublattice and for representative parameters:
\cite{Woh08} $J=0.4 t$, $\tau=0.1t$. The energy of the AO state with 
no holes, shown in Fig. \ref{strings}(a), has been used as the 
reference energy.
In agreement with the 1D nature of the three-site effective hopping
(\ref{H3sl}), the dispersion of the lowest energy state and of some
excited states shows only dependence on $k_b$.

Figure \ref{bands}(b) shows the single-particle spectral function, 
defined as
\begin{eqnarray}
\label{Ak}
A({\bf k},\omega)&=&-\frac{1}{\pi}\Im\;\sum_n \frac{Z_n( {\bf k})}{\omega -
\epsilon_{b,n}({\bf k})+ i\delta}\,,  \\
\label{Zn}
Z_n( {\bf k}) &=& |\alpha^{(n)}_{{\cal B},{\cal F}_0,{\bf k}}|^2\,,  
\end{eqnarray}
where the string ${\cal F}_0$ corresponds to the state (b) in Fig. 
\ref{strings}. In other words, this is just the weight of the bare 
hole in the wave function (\ref{psin}). As expected, the 
dispersionless bands in Fig. \ref{bands}(a) have practically no spectral 
weight --- as will be seen below, however, these bands give a dominant 
contribution to the optical conductivity.

The spectral weight of the lowest peak --- which would form the
quasiparticle band at finite doping --- shows a weak 
${\bf k}$-dependence which can be understood as follows: For momenta 
near the minimum (maximum) of the dispersion, the hole gains (looses) 
energy by propagation. Since the dominant mechanism of propagation is
the hopping of the bare hole via the three-site hopping term (\ref{H3sl})
--- see the transition between states in Fig. \ref{strings} from (b) to 
(d) --- this gain (loss) in energy will be larger if the weight of the 
bare hole in the wave function is larger. The weight of the bare hole, 
however, also gives the spectral weight of the quasiparticle peak. 
Therefore the weight of the peak is larger (smaller) near the minimum 
(maximum) of the dispersion.

\section{Optical conductivity}
\label{sec:optic}

\subsection{Numerical analysis}
\label{sec:num}

In the next step we discuss the optical conductivity, which is defined
as $\sigma_{\alpha}(\omega)=\sigma_{{\cal A},\alpha}
(\omega)+\sigma_{{\cal B},\alpha}(\omega)$, with the conductivity for
sublattice ${\cal S}\in\{{\cal A},{\cal B}\}$:
\begin{eqnarray}
\sigma_{{\cal S},\alpha} (\omega)&=&
\sum_{{\bf k}} \sum_{n=1} \;\frac{1}{\omega}\;
\left|\left\langle \Psi^{(n)}_{{\cal S},{\bf k}}\right|j_\alpha\left|
         \Psi^{(0)}_{{\cal S},{\bf k}} \right\rangle\right|^2 \nonumber \\
&\times& n_{{\cal S}{{\bf k}}}\;\delta\{\omega-
( \epsilon_{{\cal S}n}({\bf k}) - \epsilon_{{\cal S}0}({\bf k}))\}.
\label{optcon}
\end{eqnarray}
Here $\alpha\in\{ a,b \}$ denotes the direction of the current operator
$j_\alpha$, $|\Psi^{(n)}_{{\cal S},{\bf k}}\rangle$ are the approximate
single-hole eigenstates (\ref{psin}) and
$\epsilon_{{\cal S}n}({\bf k})$ are the corresponding eigenvalues.
The $n_{{\cal S}{\bf k}}$ denote the ground state occupation
numbers of these states which we assume to be different from zero
only for the lowest band labeled by $n=0$. For a given level of hole
doping $x$ they are determined by adjusting the Fermi energy. This 
implies that we are assuming that for finite hole concentration the
lowest band for each sublattice is filled according to the Pauli 
principle. This procedure is reasonable for low density of doped holes 
$x$.

We proceed to the discussion of the current operator $j_\alpha$ (for 
more clarity the index $\alpha$ is skipped below). For the
original Hamiltonian (\ref{rawh}) this is given by
\begin{equation}
 {\bf j} =i t \sum_{\delta=\pm 1}
\left(\delta \hat{\textbf b}\; 
c^{\dagger}_{i+\delta {\bf\hat{b}},a}c_{i,a}^{}+\delta\hat{\textbf a}\; 
c^{\dagger}_{i+\delta {\bf\hat{a}},b}c_{i,b}^{}\right).
 \label{curoporbdeg}
\end{equation}
At this point one has to bear in mind that the wave functions 
(\ref{psin}) are (approximate) eigenstates of the strong coupling 
Hamiltonian (\ref{t2gmodel}) rather than the original model Eq. 
(\ref{rawh}). It is well known that in order to obtain consistent 
results for a system described by the original Hamiltonian (\ref{rawh}) 
it is necessary to subject the operator in question --- here the current 
operator --- to the same canonical transformation (\ref{canonical}) as 
the Hamiltonian itself. This property has been pointed out in the strong 
coupling expansion for the
spin Hubbard model,\cite{Esk94} and we follow here this procedure.

The result of a similar calculation for the present orbital problem
is the strong coupling current operator:
\begin{eqnarray}
\label{trancurr}
j_{t_{2g}}\! &=&{\cal P} \left( j_{t}+j_{\rm 3s}^{(l)} +j_{\rm
3s}^{(d)}\right){\cal P}+O\left(\frac{t^3}{U^2}\right),\\
\label{jt}
j_{t}\!&=&-i t \sum_{ i}
\left({\bf \hat{a}}\; {c}^{\dagger}_{i,b}{c}^{}_{i+{\bf \hat{a}},b} +
{\bf \hat{b}} \; {c}^{\dagger}_{i,a}{c}^{}_{i+{\bf \hat{b}},a}
-\mbox{H.c.}\right), \\
\label{j3sl}
j_{\rm 3s}^{(l)}\! &=& -2 {\bf \hat{a}} \;i \tau \sum_{i}
\left({c}^\dag_{i-{\bf \hat{a}},b}{n}^{}_{i,a}
{c}^{}_{i+{\bf \hat{a}},b}-\mbox{H.c.}\right) \nonumber\\
&-&2 {\bf \hat{b}} \;i\tau\sum_{i}\left({c}^\dag_{i-{\bf\hat{b}},a}
{n}^{}_{i,b}{c}^{}_{i+{\bf \hat{b}},a}-\mbox{H.c.}\right),\\
\label{j3sd}
j_{\rm 3s}^{(d)}\! &=&\mp({\bf \hat{a}}-  {\bf \hat{b}})i\tau \sum_{i}
\left({c}^\dag_{i\pm{\bf \hat{b}},a}{c}^{}_{i,a}
{c}^\dag_{i,b}{c}^{}_{i\pm{\bf \hat{a}},b}-\mbox{H.c.}\right) \nonumber\\
&\mp&\!\!\!({\bf \hat{a}} + {\bf \hat{b}})i\tau \sum_{i}
\left({c}^\dag_{i\mp{\hat{b}},a}{c}^{}_{i,a}
{c}^\dag_{i,b}{c}^{}_{i\pm{\bf \hat{a}},b} -\mbox{H.c.}\right).
\end{eqnarray}
It is instructive to trace back the origin of nontrivial terms appearing in 
Eq. (\ref{trancurr}). The terms which are of order $t^2/U$ stem from
processes during which double occupancies are created at the
intermediate stage due to the NN hopping  in Eq. (\ref{curoporbdeg}) and
later removed by terms in the operator ${\cal U}$ which are of order $t/U$
or vice versa. It turns out that the process during which an electron
with $b$ orbital flavor moves from one site to its neighbor and back, see
sites $i$ and $j$ in Fig. \ref{strings}(a), does not contribute to the
transformed current operator $j_{t_{2g}}$ due to the cancelation which 
originates from the sign dependence of the prefactor in Eq. 
(\ref{curoporbdeg}) related to the hopping direction. Since there is no 
such dependence in the case of hopping term in the Hamiltonian (\ref{Ht2g}), 
the mentioned process gives rise to the exchange term (\ref{HJ}) in the 
effective Hamiltonian.

The above cancelation does not take place in the processes with transfer
the hole by two lattice spacings. In the first process $b$ orbital moves
horizontally, as from site $m$ to site $l$ in Fig. \ref{strings}(b),
followed by another hop in the same direction onto an empty (hole) site
--- this gives rise to the  contribution (\ref{j3sl}) to the effective
current. In a second process, another $b$ orbital moves horizontally,
and another hole is created along the diagonal of a plaquette (not
shown). Such processes give rise to the contribution (\ref{j3sd}) to the
effective current. Now, in the same way as for terms in the current
operator related with NN hopping (\ref{jt}), we may deduce the matrix
element of the current operator for terms related with further hopping
in the direction ${\hat{\textbf a}}$ (or ${\hat{\textbf b}}$). For the
pairs of states shown in  Figs. \ref{strings}(b), \ref{strings}(d) and in
Figs. \ref{strings}(b), \ref{strings}(e), one finds the matrix elements
$-2i\tau$ ($0$) and $i\tau$ ($-i\tau$), respectively.

The transformed current operator (\ref{trancurr}) is next used to
compute matrix elements between string states. For example, we find that 
the matrix element for the ${\hat{\textbf b}}$ component of the current 
operator $j_b$ (\ref{trancurr})
between states shown in Fig. \ref{strings}(b) and \ref{strings}(c) is $it$.
The hole shift occurs downwards. The overall sign of the matrix term is
positive, as the string states (b) $c_{i,b}|{\rm AO}\rangle$, and
(c) $c^{}_{j,a}c^{\dag}_{i,a}c^{}_{i,b}|{\rm AO}\rangle$, are defined
in the hole language which brings about an additional sign change with
respect to the prefactor appearing in Eq. (\ref{curoporbdeg}).
The matrix element of $j_a$ for the same pair of states vanishes.

\begin{figure}[t!]
 \centering
\includegraphics[width=8.2cm]{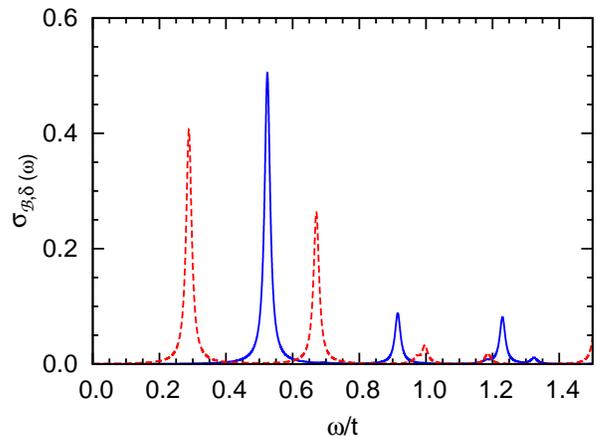}
\caption{(Color online) Contribution to the optical conductivity
$\sigma_{{\cal B}\delta}(\omega)$ measured along the
$\delta={\hat{\textbf a}}$  (${ \hat{\textbf b}}$) direction (solid line)
and along the ${\hat{\textbf b}}$ (${\hat{\textbf a}}$) direction
(dashed line), as obtained from transitions between states propagating
along the ${\hat{\textbf a}}$ (${\hat{\textbf b}}$) direction
at hole doping $x=0.1$.
The total spectrum involving transitions from both kind of states is the
sum of both contributions.
Parameters: $J=0.4t$, $\tau=0.1t$.
Lorentzian broadening of width $0.01 t$ has been used.}
\label{sab}
\end{figure}

The spectra of optical conductivity (\ref{optcon}) (obtained by
applying the Lorentzian broadening  width of $0.01 t$)
are presented in Fig. \ref{sab}.
The solid line depicts the contribution to the conductivity
measured along the $\delta={\hat{\textbf a}}$ direction from
transitions between states propagating along  the same direction,
while the dashed line depicts the conductivity measured along the
$\delta={\hat{\textbf b}}$ direction. The true response is the mixture
of contributions from states propagating in both directions and for both
$\delta={\hat{\textbf a}}$ and $\delta={\hat{\textbf b}}$. It is given
by the superposition of the spectra plotted spectra in Fig. \ref{sab}. 

\subsection{Physical picture of transitions}
\label{sec:picture}

Next we give a brief discussion of the physical significance of the
optical transitions. In the discussion of hole motion we have seen
that the dominant hopping term ${\cal H}_t$ does not lead to the
coherent propagation of a hole because it creates a string of
defects in the orbital order whence the energy increases linearly
with the number of hops, i.e., with the string length. Coherent 
propagation is enabled only by the
conditional NNN hopping term (\ref{H3sl}). Let us assume for the moment
that this term is switched off. Then we can think of localized
eigenstates of the remaining Hamiltonian
\begin{equation}
|\Psi_{i,\nu}\rangle =
\sum_{{\cal F}} \;\alpha^{(\nu)}_{{\cal F}}
|\Psi_{i,{\cal F}} \rangle.
\label{local}
\end{equation}
There are several symmetry operations which transform the
states $|\Psi_{i,\nu} \rangle$ into one another:
inversion, rotation by $\pi$ and reflection by the $\hat{b}$ ($\hat{a}$)
axis for $i\in {\cal A}$ ($i\in {\cal B}$). This corresponds to the
symmetry group $C_{2v}$ and the local eigenstates
(\ref{local}) accordingly realize irreducible
representations of this group.
Next, the Bloch states (\ref{psin}) may alternatively be written as
follows,
\begin{equation}
\left|\Psi^{(n)}_{{\cal S},{\bf k}}\right\rangle=
\sum_{\nu} \; c_{\nu}^{(n)} |\Psi_{{\bf k},\nu}\rangle,
\end{equation}
using the short-hand notation,
\begin{equation}
|\Psi_{{\bf k},\nu}\rangle =
\sqrt{\frac{2}{N}}\sum_{j}\;e^{i {\bf k} {\bf R}_j}\;
T_{{\bf R}_j} |\Psi_{i,\nu}\rangle. 
\end{equation}
This formulation --- which is completely analogous to a local 
combination of atomic orbitals (LCAO) ansatz comprising $s$-like, 
$p$-like, $d$-like basis functions, {\it etcetera\/} --- 
immediately clarifies the nature of the optical transitions: 
these are simply dipole-like transitions between the approximate local 
eigenstates $|\Psi_{i,\nu}\rangle$ generated by the interplay of 
hopping term ${\cal H}_t$ and the `string potential'. Since the current 
operator is e.g. odd under rotation by $\pi$, a matrix element 
$\langle\Psi_{i,\mu} | j_\alpha | \Psi_{i,\nu}\rangle$
is different from zero only if the two states, 
$|\Psi_{i,\mu}\rangle$ and $|\Psi_{i,\nu}\rangle$ have opposite parity. 
Since, moreover, the band with the lowest energy, $n=0$ in Eq. 
(\ref{psin}), has the totally symmetric ground state of the local 
Hamiltonian as its largest component, the peaks in the optical spectra 
give essentially --- with only a small broadening due to weak
dispersion --- by the excitation energies of the states 
with odd parity. A very similar interpretation was also
given\cite{Ede95,Voijta_Becker,Plakida} for the {\it `mid infrared'}
spectral weight observed in numerical studies of the spin
$t$-$J$ model\cite{Ste90} and applies possibly to
cuprate superconductors.

\section{Summary and conclusions}
\label{sec:summa}

We have investigated the optical conductivity in the 2D orbital model 
as a generic model for studying orbital polarons in doped orbitally 
ordered ($t_{2g}$ or $p$) systems and capturing the essential physics.
That effectively spinless model is applicable to planar systems like
some transition metal oxides when (due to the tetragonal crystal field)
degenerate $yz$ and $xz$ orbitals are active and singly occupied in a
ferromagnetically ordered plane, or to cold atom systems in optical
lattices when $p$ orbitals are active.
It is demonstrated that in the presence of strong electron
correlations, the tendency towards confinement determines the
properties of the model, i.e., it can be viewed as a system of
weakly coupled potential wells. The potential well physics originates
from exchange energy increase induced by the sequences of orbitals
flipped by a hole introduced by doping, when it moves in the
orbitally ordered Mott insulator and generates a string potential
along its path.

We have shown that propagating bands are much narrower in the
present orbitally ordered system than in the spin $t$-$J$ model.
Results of the analysis based on a variant of exact diagonalization,
motivated by the string picture, suggest the formation of bands in the
optical spectrum within the Mott-Hubbard gap, by analogy with the 
mid-infrared band observed in doped cuprates.\cite{Phi10} We predict 
that the optical conductivity would have this form in weakly doped Mott
insulators with active orbital degrees of freedom, while at higher 
doping orbital stripes would form.\cite{Wro10} Similar to the spin 
dynamics of stripes in superconducting cuprates,\cite{Kru03} one expects 
qualitative changes in the optical conductivity for systems with domains 
of AO order separated by orbital stripes, which is an interesting topic 
for future studies. Other challenges are posed by 
orbital superfluidity in the $p$-band of a bipartite optical square 
lattice investigated recently,\cite{Wir11} or by spin-orbital systems,
where an orbiton may separate from a spinon and propagate through a 
lattice as a distinct quasiparticle.\cite{Sch12}

\acknowledgments

We thank Peter Horsch for insightful discussions and comments.
A.~M.~Ole\'s acknowledges support by the Polish National
Science Center (NCN) Project No. N202~069639.

\appendix*

\section{Origin of optical transitions}
\label{sec:appa}

The aim of this section is to analyze the mechanism underlying the 
doping induced formation of states lying within the Mott-Hubbard gap
of a correlated insulator which is orbitally ordered. Furthermore,
it will be discussed, how the structure of those states  influences
the optical response. We believe that despite some simplifications,
conclusions which will be drawn from this analysis are applicable
to more complex situations encountered in real systems revealing 
orbital order.

Due to the simplicity of the analyzed model the relevant part of
the Hilbert space consists of states which can be obtained  by
subsequently shifting the hole created in the AO state, the situation
shown in Fig. \ref{strings}(b), to nearby sites. Thus we could use that
original position of the hole --- $\tilde{\bf R}({\bf R},{\cal F})$
[here $\tilde{\bf R}({\bf R},{\cal F})$ stands for the function of
the present hole position ${\bf R}$ and the set of sites on which
orbitals have been flipped (${\cal F}$)] to label a given string
state $|\Psi_{{\bf R},{\cal F}} \rangle$. In order to determine
$\tilde{\bf R}({\bf R},{\cal F})$  in a unique way, we additionally
demand that the total length of hole path necessary to create the state
$|\Psi_{{\bf R},{\cal F}} \rangle$ from a state representing a hole
created at the site $\tilde{\bf R}({\bf R},{\cal F})$ in the AO state is
minimal. For example, in the case of the state shown in Fig.
\ref{strings}(f) the vector ${\bf R}$ refers to the site $n$ while the
vector $\tilde{\bf R}({\bf R},{\cal F})$ to the site $i$. In principle
there could be more than one "original site'' obeying those conditions
for a given pair $\{{\bf R}$, ${\cal F}\}$. In that case we would
arbitrarily choose a single $\tilde{\bf R}({\bf R},{\cal F})$ and the
new method of state labeling by the new pair
$\{\tilde{\bf R}({\bf R},{\cal F}), {\cal F}\}$ would also work.

\begin{figure}[t!]
 \centering
\includegraphics[width=8.4cm]{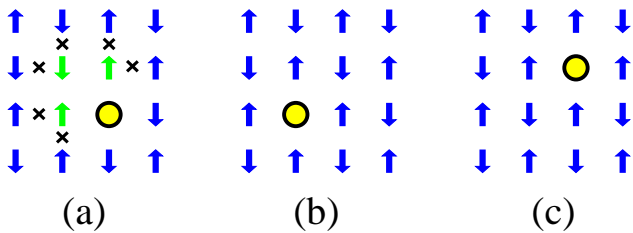}
\caption{(Color online)
The mechanism of the Trugman process\cite{Trugman} allowing for hole
deconfinement in a doped antiferromagnet described by the Ising model.
(a) A hole doped to the Mott insulator with the antiferromagnetic order
(spins are represented by arrows) generates six bonds with parallel
spins by interchanging its position with spins on a plaquette after 
three hops.
By a clockwise/anticlockwise hopping by three more steps one arrives at
states (c)/(b), where all spin excitations are removed and the 
antiferromagnetic state is repaired --- hence such processes couple the 
configurations (b) and (c). Broken bonds that cost the classical energy 
$\frac12 J$ each are marked by~$\times$.}
\label{trugman}
\end{figure}

The possible ambiguity in determining the original position of the
hole for a given string state would give rise to a new channel of
coherent hole propagation which is known in the case of doped
antiferromagnets. Fig. \ref{trugman} depicts the so-called Trugman
process \cite{Trugman} which brings about hole deconfinement on a square
lattice even in systems with Ising-type anisotropic exchange interaction
as in the present orbital $t$-$J_z$ model. The state shown in Fig. 
\ref{trugman}(a) can be obtained both by shifting the hole clockwise or 
anticlockwise around the elementary plaquette by three lattice spacings 
(initially the hole replaced a $\downarrow$-spin as shown in Figs. 
\ref{trugman}(b) and \ref{trugman}(c), respectively).
In other words, by performing one and a half of the circular movement of
the hole on a given plaquette, the end effect is that the hole has moved
along the plaquette diagonal without bringing about spin flips in the 
N\'eel state, which gives rise to the weak coherent hole propagation.

Now, with the help of the computer algebra, we will demonstrate that in
the present two-orbital problem the ambiguity in determining the
original position of the hole for a given string state does not exist 
in the case of a string state generated by a single hole, with a small
number of twisted orbitals in the $|{\rm AO}\rangle$ state. Thus, it 
seems that the only channel allowing for hole propagation is due to the 
Hamiltonian induced coupling between states representing a hole created 
in the $|{\rm AO}\rangle$ state, as
the states depicted in Figs. \ref{strings}(b) and \ref{strings}(d). This
kind of hopping is mediated by the small $\tau \ll t$ term (\ref{H3sl}).

We proceed now to provide the justification of that statement when
restricted to the low energy sector of the Hilbert space, i.e.,
consisting of states with limited number of flipped orbitals in the
$|{\rm AO}\rangle$ state. The Hamiltonian matrix represented in terms 
of states $|\Psi_{{\cal S},{\cal F},{\bf k}} \rangle$ consists of two 
decoupled blocks. The first (second) block is formed by states coupled 
with the state representing a hole created in the perfect 
$|{\rm AO}\rangle$ state at a
site belonging to the sublattice $\cal A$ ($\cal B$). Despite that both
blocks show explicit dependence on both $\{k_a,k_b\}$, the energies of
eigenstates for each block disperse along a single direction, either
(10) or (01) in the 2D Brillouin zone.

By representing the Hamiltonian matrix in terms of
\begin{equation}
|\tilde{\Psi}_{{\cal S},{\cal F},{\bf k}} \rangle =
\sqrt{\frac{2}{N}} \sum_{i \in {\cal S}} e^{i {\bf k}
\tilde{\bf R}({\bf R}_i,{\cal F})}
|\Psi_{{\bf R}_i,{\cal F}} \rangle   \label{newpropstat},
\end{equation}
which is equivalent with performing a kind of gauge
transformation, we get rid of the superfluous momentum dependence
of the Hamiltonian matrix.  Furthermore, if we neglect the term
(\ref{H3sl}), the Hamiltonian matrix lacks any dependence on momentum,
which shows that the hole deconfinement occurs here solely due to
that term and that the hole becomes confined when $\tau$ is set to zero.
Those findings are restricted to the basis considered by us which is
limited by the path length. Furthermore, we do not
probe the part of the Hilbert space which is not coupled by a
Hamiltonian power to original states, representing a hole created in
the $|{\rm AO}\rangle$  state. Thus, a general mathematical proof of 
the hypothesis regarding the nature of the propagation and valid for 
the whole Hilbert space is still needed.

To identify the origin of optical transitions contributing
to the spectrum depicted in Fig. \ref{sab}, we analyze now the
properties of the Hamiltonian matrix represented in terms of
states (\ref{newpropstat}) [i.e., in the case when the original
hole position is used to label string states], when $\tau$ is set
to $0$, which means that only the ``fast'' part of the Hamiltonian
is considered. It turns out that this part determines the overall
structure of the energy-band hierarchy. Due to the lack of energy
dispersion the bands are now completely flat, but  their positions
correspond very well to the sequence of bands obtained for finite
$\tau$ and shown in Fig. \ref{bands}. The picture which emerges
from that correspondence is that the physics of hole hopping in the 
orbitally ordered background at large energy scale is determined by 
fast moves to NN sites, accompanied by the creation of defects in the 
orbital arrangement. Due to the increasing length of defect sequences  
(strings) a hole behaves like a particle in a potential well. 
Consequently, band hierarchy corresponds to the sequence of 
eigenenergies for the corresponding problem of the particle in the well. 
On the other hand, the energy dispersion is determined by ``slow'' 
hopping which brings about the modification at the energy scale 
$\sim \tau \ll t$ and can be viewed as a perturbation introduced on top 
of the robust structure of eigenstates arising for the potential-well
problem. For finite $\tau$, the Hamiltonian matrix block
formed by states coupled with the state representing a hole removed
from the  $\cal A$ ($\cal B$) sublattice in the $|{\rm AO}\rangle$ state 
does not show any dependence on $k_a$ ($k_b$), which explicitly
demonstrates that the hole propagation is 1D.

\begin{figure}[t!]
 \centering
\includegraphics[width=8.4cm]{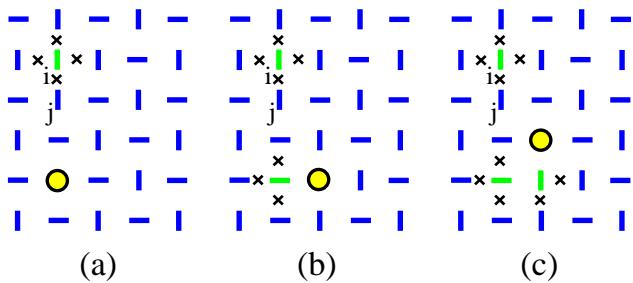}
\caption{(Color online)
Artist's view of some disconnected string states which are not coupled
by NN hopping $t$ with the original state of a hole doped in the AO state
shown in Fig. \ref{strings}(b). After moving by one hopping process $t$
from site $i$ to site $j$, the hole hops downwards by the effective 
hopping $\tau$ and leaves behind an island of excited bonds (a). Further 
hopping of the hole by either (b) one $t$ step, or (c) two $t$ steps, 
generates more defects in the $|{\rm AO}\rangle$ state. Broken bonds 
($\times$) as in Fig. 4.  }
\label{stringsymm}
\end{figure}

The point group of the underlying orbital background is $C_{2v}$.
The Brillouin zone gets folded due to the staggered form of
the AO order and all parameters like the energy are periodic with
the periodicity $(\pi,\pi)$ and $(\pi,-\pi)$. The bottom of the lowest
energy band  lies on the lines $(\pi/2,k_b)$  and $(k_a,\pi/2)$
(and on lines equivalent by symmetry) for two orthogonal directions of 
the 1D hole propagation, respectively. Within the single particle 
approximation which we apply here, the Mott insulator doped with holes 
starts to fill orbital polaron bands near their bottoms. Even for 
nonzero $\tau$ the Hamiltonian blocks represented in the basis of states
(\ref{newpropstat}) are fully symmetric with respect to the point
group $C_{2v}$ for wave vectors lying at the band bottom. For
example, the block in the sector consisting of states lacking
dispersion in the $\hat{\textbf b}$ direction is symmetric at the
wave vector $(\pi/2,k_b)$ with respect to the reflection in the $a$
axis, as it lacks the dependence on $k_b$. The inversion in the $b$
axis transforms  $(\pi/2,k_b)$ into $(-\pi/2,k_b)$, which is the same 
as $(\pi/2,k_b+\pi)$ due to Brillouin zone folding, and equivalent to
$(\pi/2,-k_b)$ due to the lack of dependence on $k_b$. Thus we can
use $C_{2v}$ to classify the symmetry properties of states at the
band bottom.\cite{class}

The ground state for both nonzero $\tau$ (system with propagating holes)
and for $\tau$ set to zero (system with confined holes) is even with 
respect to both reflections: in the $a$ axis ($s_a$), and in the $b$ 
axis ($s_b$). In the latter case the states like the ones depicted in 
Figs. \ref{stringsymm}(a)-\ref{stringsymm}(c) which are not coupled by a 
product of the $t$-hopping terms (\ref{Ht}) with the original state, 
with a hole created in the $|{\rm AO}\rangle$ state at site $i$ and 
shown in Fig. \ref{strings}(b), do not contribute to the ground state, 
while in the former case their weight is small. The state depicted in 
Fig. \ref{stringsymm}(a) is coupled with the state shown in Fig.
\ref{strings}(c) by the $\tau$-hopping term (\ref{H3sl}) which, on
the other hand, is coupled with the original state presented in Fig.
\ref{strings}(b) by the $t$-hopping term (\ref{Ht}). The states 
shown in Figs. \ref{stringsymm}(b) and \ref{stringsymm}(c) are both 
coupled by hopping terms with the state presented in 
Fig. \ref{stringsymm}(a).

The first excited state in the sector corresponding to the propagation
in the $\hat{\textbf a}$ direction is odd with respect to the reflection 
$s_a$ and even with respect to $s_b$. Since the current operator 
(\ref{trancurr}) is an axial vector, its $b$ component (which is odd 
with respect to $s_a$) couples the first excited state with the fully 
symmetric ground state, which gives rise to the contribution to the 
optical weight in the form of the first peak from the left shown in Fig. 
\ref{sab}. Upon doping, the optical transitions occur at momenta located 
in the vicinity of the band minimum, which explains why the peak gets a 
finite width. In the system with no propagating holes, the states such 
as those depicted in Fig. \ref{stringsymm}, which are not coupled by NN 
hopping to the original state, do not contribute to the first excited 
state, while their weight is small when the holes propagate.



\begin{thebibliography}{99}
\bibitem{noteu} This condition is fulfilled in transition metal oxides;
                here an effective interaction $U$ includes Hund's exchange
                for the high-spin state, see also Ref. \onlinecite{Woh08}.
\bibitem{class} Strictly speaking, since the element $O$ of the point
                group transforms the states
                $|\Psi_{{\cal S},{\cal F},{\bf k}} \rangle$ and
                $|\tilde{\Psi}_{{\cal S},{\cal F},{\bf k}}\rangle$ into
                $|\Psi_{{\cal S},O{\cal F},O{\bf k}} \rangle$ and
                $|\tilde{\Psi}_{{\cal S},O{\cal F},O{\bf k}}\rangle$,
                respectively, in the classification for the wavevector
                $(\pi/2,k_b)$ and $(k_a,\pi/2)$ we use
                $|\tilde{\Psi}_{{\cal S},{\cal F},k_a} \rangle=
                \sum_{k_b} |\tilde{\Psi}_{{\cal S},{\cal F},{\bf k}}\rangle$
                and $|\tilde{\Psi}_{{\cal S},{\cal F},k_b} \rangle=
                \sum_{k_a} |\tilde{\Psi}_{{\cal S},{\cal F},{\bf k}}\rangle$;
                these states have the same matrix elements in the respective
                block for the Hamiltonian and the current operator as any
                term in the sum.
\end{thebibliography}
\end{document}